\documentclass[12pt]{article}
\usepackage{times}
\usepackage{geometry}
\usepackage{graphicx}	
\usepackage[utf8]{inputenc}
\usepackage{enumitem,amssymb}
\usepackage{wrapfig}
\usepackage{ragged2e}
\newlist{thematic}{itemize}{8}
\setlist[thematic]{label=$\square$}
\usepackage{pifont}
\usepackage{xspace}
\newcommand{\cmark}{\ding{51}}%
\newcommand{\done}{\rlap{$\square$}{\raisebox{2pt}{\large\hspace{1pt}\cmark}}%
\hspace{-2.5pt}}

\newcommand{\tcm}{21$\,$cm\xspace}  

\newcommand{\LCDM}{$\Lambda$CDM\xspace}
\usepackage{titlesec}
\titleformat*{\section}{\large\bfseries}

\begin{document}
\raggedright
\huge
Astro2020 Science White Paper \linebreak

Cosmology with the Highly Redshifted \tcm Line \linebreak
\normalsize

\noindent \textbf{Thematic Areas:} \hspace*{60pt} $\square$ Planetary Systems \hspace*{10pt} $\square$ Star and Planet Formation \hspace*{20pt}\linebreak
$\square$ Formation and Evolution of Compact Objects \hspace*{31pt} 
\done Cosmology and Fundamental Physics \linebreak
  $\square$  Stars and Stellar Evolution \hspace*{1pt} $\square$ Resolved Stellar Populations and their Environments \hspace*{40pt} \linebreak
  $\square$    Galaxy Evolution   \hspace*{45pt} $\square$             Multi-Messenger Astronomy and Astrophysics \hspace*{65pt} \linebreak
  
\textbf{Principal Author:}

Name: Adrian Liu 
 \linebreak						
Institution: McGill University  
 \linebreak
Email: \verb|acliu@physics.mcgill.ca|
 \linebreak
Phone: (514) 716-0194
 \linebreak
 
\textbf{Co-authors:} (names and institutions)
  \linebreak James Aguirre (University of Pennsylvania), Joshua S. Dillon (UC Berkeley), Steven R. Furlanetto (UCLA), Chris Carilli (National Radio Astronomy Observatory), 
 Yacine Ali-Haimoud (New York University),
 Marcelo Alvarez (University of California, Berkeley),
 Adam Beardsley (Arizona State University),
 George Becker (University of California, Riverside),
 Judd Bowman (Arizona State University),
 Patrick Breysse (Canadian Institute for Theoretical Astrophysics),
 Volker Bromm (University of Texas at Austin),
 Philip Bull (Queen Mary University of London),
 Jack Burns (University of Colorado Boulder),
 Isabella P. Carucci (University College London),
 Xuelei Chen (National Astronomical Observatories, Chinese Academy of Sciences), 
 Tzu-Ching Chang (Jet Propulsion Laboratory),
 Hsin Chiang (McGill University),
 Joanne Cohn (University of California, Berkeley),
 David DeBoer (University of California, Berkeley),
 Olivier Dor\'{e} (Caltech/JPL),
 Cora Dvorkin (Harvard University),
 Anastasia Fialkov (Sussex University),
 Nick Gnedin (Fermilab),
 Bryna Hazelton (University of Washington),
 Jacqueline Hewitt (Massachusetts Institute of Technology), 
 Daniel Jacobs (Arizona State University),
 Kirit Karkare (University of Chicago/KICP), 
 Marc Klein Wolt (Radboud University Nijmegen),
 Saul Kohn (The Vanguard Group),
 Leon Koopmans (Kapteyn Astronomical Institute),
 Ely Kovetz (Ben-Gurion University),
 Paul La Plante (University of Pennsylvania),
 Adam Lidz (University of Pennsylvania),
 Yin-Zhe Ma (University of KwaZulu-Natal),
 Yi Mao (Tsinghua University),
 Kiyoshi Masui (Massachusetts Institute of Technology),
 Andrei Mesinger (Scuola Normale Superiore, Pisa),
 Jordan Mirocha (McGill University),
 Julian Munoz (Harvard University),
 Steven Murray (Arizona State University),
 Laura Newburgh (Yale University),
 Aaron Parsons (University of California, Berkeley),
 Jonathan Pober (Brown University),
 Jonathan Pritchard (Imperial College),
 Benjamin Saliwanchik (Yale University),
 Jonathan Sievers (McGill University),
 Nithyanandan Thyagarajan (National Radio Astronomy Observatory),
 Hy Trac (Carnegie Mellon University),
 Eli Visbal (Flatiron Institute),
 Matias Zaldarriaga (Institute for Advanced Study)
 \newpage
\textbf{Abstract  (optional):}
In addition to being a probe of Cosmic Dawn and Epoch of Reionization astrophysics, the \tcm line at $z>6$ is also a powerful way to constrain cosmology. Its power derives from several unique capabilities. First, the \tcm line is sensitive to energy injections into the intergalactic medium at high redshifts. It also increases the number of measurable modes compared to existing cosmological probes by orders of magnitude. Many of these modes are on smaller scales than are accessible via the CMB, and moreover have the advantage of being firmly in the linear regime (making them easy to model theoretically). Finally, the \tcm line provides access to redshifts prior to the formation of luminous objects. Together, these features of \tcm cosmology at $z>6$ provide multiple pathways toward precise cosmological constraints. These include the ``marginalizing out" of astrophysical effects, the utilization of redshift space distortions, the breaking of CMB degeneracies, the identification of signatures of relative velocities between baryons and dark matter, and the discovery of unexpected signs of physics beyond the \LCDM paradigm at high redshifts.

\pagebreak

In the next decade, measurements of the \tcm line of hydrogen at high redshifts have the potential to make a profound impact on cosmology and fundamental physics. In this white paper, we focus on cosmological and fundamental physics applications at $z > 6$, leaving a discussion of cosmological probes at $z < 6$ and astrophysical probes at all redshifts to Ref. \cite{Stage2}. At $z > 6$, the \tcm line provides a tracer of structure that is sensitive to cosmology in its own right, and is unique in its ability to open a discovery space for the unexpected. Alternatively, it may be used in concert with other cosmological probes by removing astrophysical ``nuisance" parameters. \tcm cosmology builds on the foundations of observational cosmology provided by galaxy surveys and the Cosmic Microwave Background (CMB) in several unique ways
\cite{tegmarkzaldarriaga2009}:

\begin{itemize}
    \item {\bf Sensitivity to energy injection.} The strength of absorption or emission in the \tcm at early times is strongly coupled to the injection of energy into the intergalactic medium (IGM) by the first stars, by accreting compact objects, by shocks, and by a variety of potential exotic processes.
    
    \item {\bf Access to redshifts before the formation of luminous objects.} The \tcm line is one of the only probes of our Universe at redshifts between recombination and the formation of the first luminous objects. Opening new redshift windows has already provided tantalizing hints of new physical phenomenology (see Section \ref{sec:Unexpected}).
    
    \item {\bf Orders of magnitude increase in number of measurable modes.}  Hydrogen is omnipresent in our Universe, allowing large volumes to be mapped in 3D, with the line-of-sight distance automatically given by the redshift of the line. The \tcm line can in principle access orders of magnitude more cosmological modes than galaxy surveys or the CMB \cite{Stage2}.
    
    \item {\bf The ability to probe small scale modes.} 
    The \tcm line can also access small scales. Unlike the CMB, small scales are not Silk damped, allowing measurements down to the Jeans scale. Additionally, at high redshifts the modes remain linear to smaller scales, simplifying theoretical analyses. Finally, the wide-field nature of low-frequency observations allow access to the largest scales.
    
\end{itemize}

Motivated by these possibilities, low-frequency radio astronomy has exhibited a renaissance in the last decade. Large interferometric facilities with unparalleled digital processing capacities, enormous collecting areas, and unprecedented sensitivities have been built and are now in operation. These include the Murchison Widefield Array (MWA; \cite{Bowman2013, Tingay2013}), the Low Frequency Array (LOFAR; \cite{LOFAR2013}), the Donald C. Backer Precision Array for Probing the Epoch of Reionization (PAPER; \cite{Parsons2010}), the Canadian Hydrogen Intensity Mapping Experiment (CHIME; \cite{Bandura2014}), the Hydrogen Intensity and Real-time Analysis eXperiment (HIRAX; \cite{Newburgh2016}), and the Hydrogen Epoch of Reionization Array (HERA; \cite{DeBoer2017}), the Owens Valley Radio Observatory Long Wavelength Array (OVRO-LWA; \cite{Eastwood2018}) and the Large-aperture Experiment to Detect the Dark Age (LEDA; \cite{Greenhill2012}).
 While a positive detection of the spatially fluctuating \tcm signal at $z > 6$ remains elusive, considerable progress has been made in the form of increasingly stringent upper limits. Additionally, single-element instruments with exquisite hardware and software calibration capabilities have been constructed, such as the Experiment to Detect the Global Epoch of Reionization Step (EDGES; \cite{BowmanRogers2010}), Shaped Antenna measurement of the background RAdio Spectrum (SARAS; \cite{Patra2015,Saurabh2018}), Sonda Cosmol\'{o}gica de las Islas para la Detecci\'{o}n de
Hidr\'{o}geno Neutro (SCI-HI; \cite{Voytek2014}), and the Probing Radio Intensity at high-Z from Marion instrument (PRI$^Z$M; \cite{Philip2018}). This has recently led to a possible detection of the global (i.e., spatially averaged) \tcm signal by EDGES \cite{EDGES2018}, which would provide hints of beyond-\LCDM  cosmology if confirmed.

$\qquad$To take advantage of the experimental capabilities these pathfinders will enable over the next decade, the key theoretical challenge to using the \tcm line for cosmology at $z > 6$ is to understand the coupling of complex astrophysical processes to the \tcm signal---such as the formation of the first stars and black holes, and the reionization of the IGM---which tend to obscure fundamental physics. At $z > 30$, these processes are relatively unimportant, but significant instrumentation and data analysis challenges need to be overcome for such measurements, and thus these epochs remain a future goal. However, important developments in the past decade, both experimental and theoretical, how now placed the community in a position where there are multiple strategies to access cosmology with the highly redshifted \tcm line.

\section{Accessing Fundamental Cosmology By Marginalizing Over Astrophysics}

\begin{figure*}
\centering
\includegraphics[width=0.45\textwidth,trim=0cm 0cm 0cm 0cm,clip]{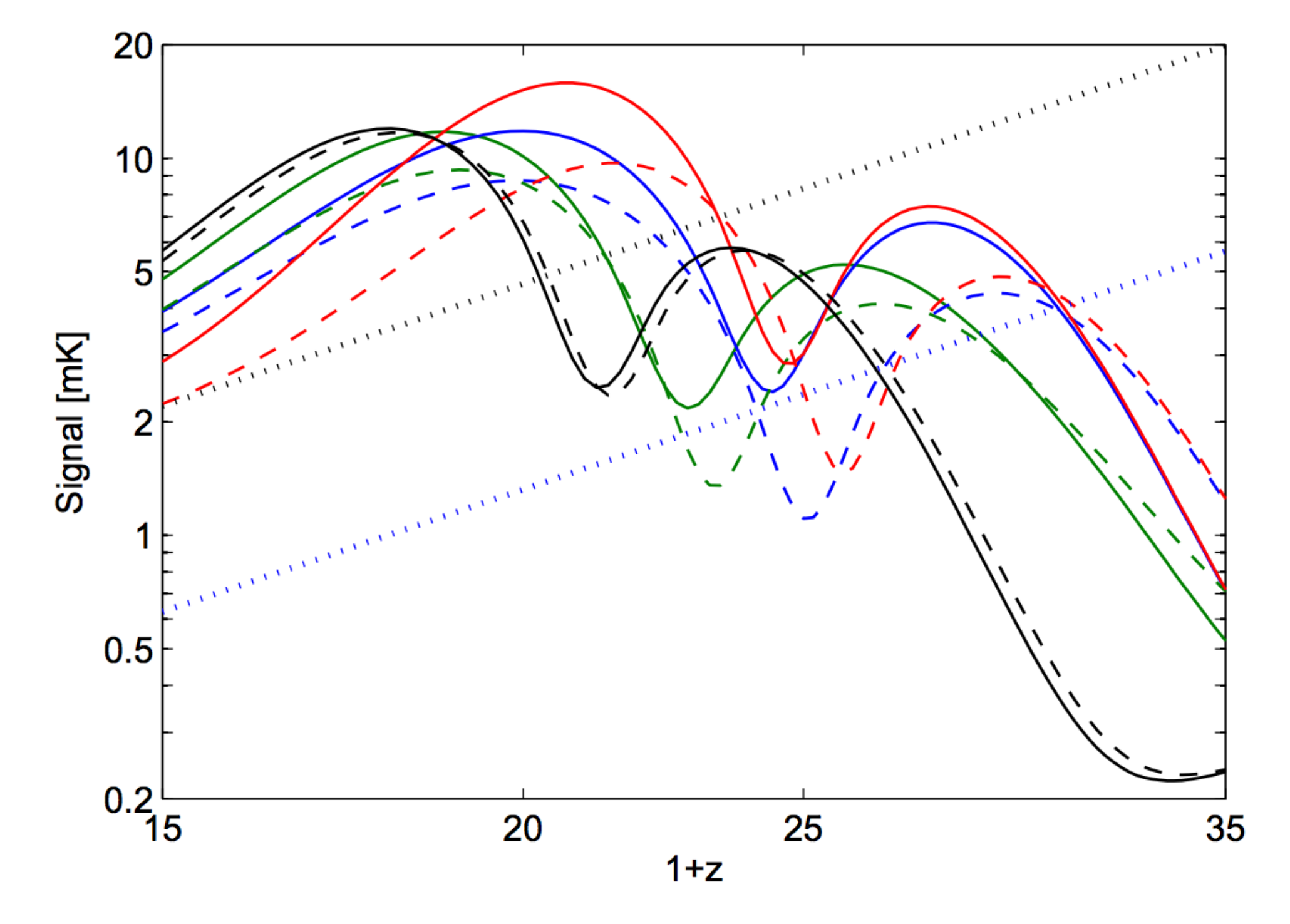}
\includegraphics[width=0.45\textwidth,trim=0cm 0cm 0cm 0cm,clip]{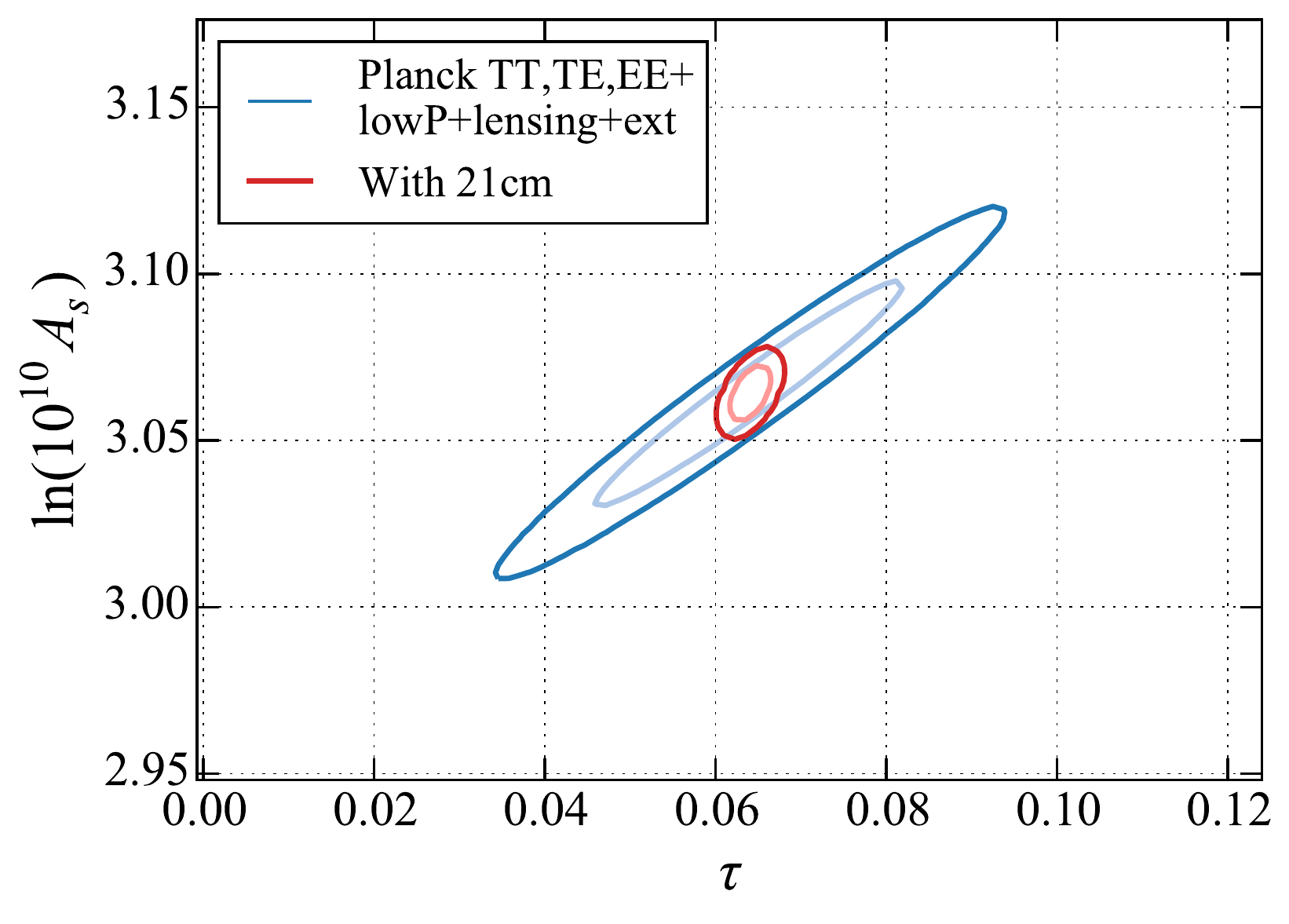}
\caption{\footnotesize {\bf Left: Relative velocity effects between baryons and dark matter have observable \tcm signatures.} Solid curves show various \tcm power spectra (evaluated at $k \sim 0.1 h/\textrm{Mpc}$) for various feedback prescriptions without relative velocity effects; dashed curve include these effects. Dotted curves show sensitivities for current-generation arrays (black) and next-generation arrays (blue).  From Ref. \cite{Fialkov2014}. {\bf Right: Using the \tcm line to place constraints on the CMB optical depth $\tau$ can potentially break existing degeneracies on fundamental cosmological parameters.} Blue contours show constraints on the $\tau$ and the amplitude of the primordial power spectrum $A_s$ from Planck ``TT + TE + EE + lowP + lensing + ext" from Ref. \cite{Planck2016Cosmo}. This illustrates the degeneracy that arises from using CMB data alone. This degeneracy is substantially mitigated when \tcm constraints on $\tau$ are included. Orange contours show the resulting forecast for HERA. From Ref. \cite{Liu2016}.}
\label{fig:both}
\end{figure*}

Although the problem of separating the astrophysics of Cosmic Dawn and reionization seems difficult at first glance, theoretical and observational advances in the last decade have highlighted multiple paths towards a solution.
%

$\qquad$The first approach to extracting cosmological constraints from $z > 6$ measurements of the \tcm line is to fit the data with models that include both astrophysics and cosmology. Extracting cosmological parameters is then a matter of marginalizing out the astrophysical parameters \cite{Clesse2012}. This can result in powerful cosmological constraints, 
for instance with forecasts suggesting measurements of the sum of the neutrino masses to within $0.006\,\textrm{eV}$ \cite{Mao2008}, which compares favorably to the uncertainty of $\sim 0.02\,\textrm{eV}$ obtainable using CMB-S4 measurements \cite{CMBS4}. 
This would be an important measurement, given that a non-zero neutrino mass is one of the few glimpses of physics beyond the Standard Model.

$\qquad$Of course, constraints arising from the aforementioned procedures will be limited by our ability to correctly account for the astrophysics of $z > 6$ in our models. Naturally, given the dearth of current observations, this can be difficult to assess. This is why cosmological constraints from the \tcm line and the CMB are complementary: the former has tighter error bars at the expense of model dependence. However, several recent theoretical advances have improved the situation. Semi-analytic codes for the \tcm power spectrum at $z > 6$ have moved towards increasingly flexible parametrizations of the underlying astrophysics, enabling conservative but robust cosmological constraints \cite{ParkMesingerGreig2019}. In addition, a recent breakthrough in the theoretical modelling of reionization has shown that, contrary to previous expectations, ionization fluctuations may be efficiently describable using perturbation theory  \cite{Hoffmann2018,McQuinnD'Aloisio2018}, enabling \tcm fluctuations to be described as a bias expansion of the matter density field, expressible with a relatively small number of free parameters. Finally, a positive detection of the fluctuating \tcm signal from current and upcoming instruments would provide an opportunity to {\it test} rather than {\it assume} our parametrizations.

$\qquad$ An alternative 
is to look at redshift space distortions, which are directly sourced by the matter density field and thus bypass the complicated astrophysics governing ionization or spin temperature fluctuations \cite{BarkanaLoeb2005,McQuinn2006}. Early theoretical forecasts of this effect were based on linear theory, but detailed non-linear simulations have recently confirmed the promise of pursuing this type of measurement, at least in the early stages of reionization (when the neutral fraction is less than $\sim 40\%$) \cite{Shapiro2013}.

$\qquad$ A final observational signature that is worth pursuing is that of supersonic relative velocities between baryons and dark matter. This has measurable consequences in a variety of contexts, such as the suppression of small-scale structures. Recognition of this phenomenon in Ref. \cite{TseliakhovichHirata2010} was a key theoretical advance in this past decade, and observational follow-up of its consequences is ongoing. The high-redshift reach of the \tcm line allows access to regimes where relative-velocity effects (see Figure \ref{fig:both}) are most apparent \cite{Dalal2010,Fialkov2014}, potentially allowing constraints on possible baryon-dark matter interactions \cite{Munoz2015}. This necessitates the construction of next-generation instruments that, like many current-generation instruments, are optimized for power spectrum measurements, but have much larger sensitivities for probing Cosmic Dawn.\linebreak

{\bf To take advantage of the opportunities in this area, the following key advances are necessary:}

\begin{itemize}
    \item Existing instruments must detect and characterize the Epoch of Reionization \tcm power spectrum to high significance. This will enable astrophysical models to be tested, rather than assumed, which is a first step towards determining whether astrophysical effects can be separated from cosmological ones.
    
    \item The design and construction of next-generation interferometric arrays that can characterize the $13 < z < 30$ power spectrum to high significance, which current-generation instruments may detect only marginally.
    
\end{itemize}

\section{Accessing Fundamental Cosmology in Conjunction with the CMB}

Another way to sidestep Cosmic Dawn and reionization astrophysics is to produce joint cosmological constraints with the CMB. In particular, \tcm observations can be used to make predictions for the optical depth to the CMB, $\tau$, which is a crucial ``nuisance" parameter in CMB constraints \cite{Liu2016,FialkovLoeb2016}. This is expected to be a limiting factor in many next-generation CMB measurements taking place over the next decade \cite{CMBS4}. An independent constraint on $\tau$ from \tcm cosmology enables tighter CMB measurements on the sum of the neutrino masses $\sum m_\nu$ (to within $\sim 12\,\textrm{meV}$) or the amplitude of the primordial power spectrum $A_s$ (to better than a percent; see Figure \ref{fig:both}) \cite{Liu2016}. Note that CMB constraints on $\sum m_\nu$ are weakened if it is necessary to extend \LCDM\ beyond 6 parameters, increasing the importance of independent constraints \cite{Allison2015}.


$\qquad$ A potential weakness to this approach is that \tcm instruments do not directly measure the optical depth. Thus, \tcm-based estimates of $\tau$, while valuable, will necessarily be model-dependent. High-significance detections of reionization to first test the underlying assumptions of our models are needed, as well as approaches to developing more model-independent estimates of the ionization fraction from observational data.

\section{Accessing Fundamental Cosmology Via Unexpected Results: The EDGES detection}
\label{sec:Unexpected}

In addition to the aforementioned probes of standard cosmology, pushing the redshift frontier of \tcm cosmology enables the possibility of discovering \emph{nonstandard} cosmology.

$\qquad$ Consider the Experiment to Detect the Global EoR Signature (EDGES). While it remains unconfirmed, EDGES recently announced a tentative detection of the \emph{global} \tcm signal, i.e., the angularly averaged all-sky signal \cite{EDGES2018}. If the EDGES detection is confirmed by other experiments, it would represent the first direct observational constraints at $z \sim 17$. The EDGES signal is remarkable because its detailed shape was unexpected. The timing and narrowness of the measured absorption trough implies that star formation rates must evolve much more rapidly at $z> 10$ than expected \cite{MirochaFurlanetto2019}. Moreover, the amplitude of the trough is approximately a factor of 2 larger than is allowable under \LCDM unless there exists a previously unknown population of radio loud sources at high redshifts \cite{FengHolder2018,Ewall-Wice2018,Sharma2018,Jana2019,Fialkov2019}. This has generated a large number of theoretical interpretations, including those that involve exotic new physics such as dark photons \cite{Jia2019}, quark nuggets \cite{Lawson2018}, charge sequestration \cite{Falkowski2018}, dark matter annihilations \cite{cheung2019}, dark matter-baryon scattering \cite{SlatyerWu2018,HiranoBromm2018,Barkana2018}, interacting dark energy \cite{Costa2018}, axions \cite{Moroi2018}, relic neutrino decays \cite{Chianese2019}, and partially charged dark matter \cite{Berlin2018,Kovetz2018,Munoz2018a}. In addition, if taken at face value, the signal places competitive limits on warm dark matter \cite{Safarzadeh2018,Schneider2018}, ultralight dark matter \cite{KovetzFuzzy2018}, primordial black hole dark matter \cite{Clark2018,Hektor2018}, dark matter decay \cite{Mitridate2018}, and the primordial matter power spectrum \cite{Yoshiura2018}.

$\qquad$ The aforementioned results demonstrate the power of the newly accessible redshifts with \tcm cosmology: either we will encounter unexpected results due to new physics, or we will obtain tighter constraints on existing models. Both have the potential to indicate which extensions or revisions of \LCDM\ are necessary.\linebreak

{\bf To take advantage of the opportunities in this area, the following key advances are necessary:}

\begin{itemize}
    \item The EDGES result must be confirmed or refuted by some combination of global \tcm signal experiments and spatial fluctuation measurements at the same redshifts. Both methods will likely be required, not only because each type of measurement has its own systematic uncertainties, but because power spectrum measurements may be needed to break degeneracies between competing global signal models \cite{Fialkov2019} and to further elucidate the nature of possible anomalous global observations \cite{Munoz2018b}.
    
    \item Theoretical advances in frameworks for dark matter beyond the WIMP are needed. Many explanations invoked for the anomalous EDGES signal involve dark matter. This focus on new dark matter models is particularly timely given that direct detection dark matter experiments are approaching design sensitivities without convincing detections, as well as the continued lack of evidence for supersymmetry at the LHC.
    
\end{itemize}

\section{Conclusion: A Stepping Stone Towards the Dark Ages}

From a purely theoretical standpoint, the most attractive way to access cosmology using the \tcm line is to push to the extremely high redshift regime corresponding to the Dark Ages. During the Dark Ages, the first stars have yet to form, and thus the \tcm brightness temperature field is expected to trace density fluctuations and related quantities such as their velocities. With the large number of cosmological modes in the linear regime at these redshifts, plus the absence of Silk damping, exquisite cosmological constraints are in principle possible. For details, see the discussion in the white paper titled \emph{Fundamental Cosmology in the Dark Ages with 21-cm Line Fluctuations} submitted by Furlanetto et al. \linebreak 

{\bf To take advantage of the opportunities in this area, the following key advances are necessary:}
\begin{itemize}
    \item \tcm measurements of Cosmic Dawn and reionization are a crucial testbed for hardware and software development. Continued investment in this area therefore serves as a stepping stone towards the Dark Ages.
    
    \item The push to higher redshifts requires pushing to lower frequencies, where the dynamic ionosphere has an increasing influence on precision measurements. Detailed empirical studies of ionospheric fluctuations need to be performed, in order to inform calibration strategies that may be affected by such fluctuations. Promising ground-based observations should be pursued to ascertain precisely where the low-frequency cutoff of the ionosphere is in practice, informing possible space-based facilities in the future.
   
\end{itemize}

In conclusion, the \tcm line is a promising way to probe both astrophysics and cosmology, given the unprecedented ranges that it can reach in both redshift and scale. To access cosmology, it is necessary to model or avoid the astrophysical effects. While both approaches are challenging at $z >6$, the potentially great rewards make them worth pursuing.

\pagebreak

\bibliographystyle{abbrv}
\bibliography{biblio}

\end{document}